\documentclass[12pt]{article}
\begin{document}

\begin{center}
$\mathbf{A}$ $\mathbf{\;MODIFIED}$ $\mathbf{\;VARIATIONAL}$ $\mathbf{%
\;PRINCIPLE}$ $\mathbf{\;IN\;}$

$\mathbf{RELATIVISTIC}$ $\mathbf{\;HYDRODYNAMICS}$

\bf{II. Variations of the vector field and the projection tensor in the
general case and under definite assumptions }

B.G. Dimitrov

{{Joint Institute for Nuclear Research, Laboratory for Theoretical
Physics} }

{{Dubna 141 980, Russia} }

{E-mail: bogdan@thsun1.jinr.ru }

{\bf{Abstract}}
\end{center}

The purpose of the paper is to develop further a projection variational
approach in relativistic hydrodynamics. The approach, previously proposed in
[gr-qc/9908032], is based on the variation of the vector field and the
projection tensor (instead of the given metric tensor) and their first
partial derivatives. The previously proved property of non-commutativity of
the variation and the partial derivative in respect to the projection tensor
has been used to find all the variations. Subsequently, motivated by some
analogy with the well-known (3+1) ADM\ projection formalism, an assumption
has been made about a zero-covariant derivative of the projection tensor in
respect to the projection connection. The combination of the equations for
the variations of the projective tensor with covariant and contravariant
indices has lead to the derivation of an important and concisely written
relation: the derivative of the vector field length is equal to the
''twice'' projected along the vector field initial Christoffell connection.
The result is of interest due to the following reasons: 1. It is a more
general one and contains in itself a well-known formulae in affine
differential geometry for the so called equiaffine connections (admitting
covariantly conserved tensor fields), for which the trace of the connection
is equal to the gradient of the logarithm of the vector field length. 2. The
additional term is the projected (with the projection tensor) initially
given connection and accounts for the influence of the reference system on
the change of the vector field's length, measured in this system. 3. The
formulae has been obtained within the proposed formalism of non-commuting
variation and partial derivative.

\begin{center}
\bf{{I. INTRODUCTION.} }
\end{center}

A key object in relativistic hydrodynamics is the energy -momentum tensor,
which by the nature of its physical foundations should embody in itself not
only the properties of the gravitational field, but also the mechanical
properties of matter (in the monograph [1] called also ''a continuum'' and
having a more concrete meaning - matter should not be described by means of
a discrete model). For example, if according to the ''continuous'' model to
each particle of the matter a four-velocity vector $u_{i{ }}$can be
assigned, then the energy-momentum tensor is defined as

\begin{equation}
T_{ij}\equiv \mu u_iu_j-S_{ij},  \label{1}
\end{equation}
where $\mu $ is the proper density of energy (or mass) and $S_{ij}$ is the
{stress tensor, }accounting for the action of tension (or surface
forces). The idea that such tension forces can contribute to the energy of
the system ''gravitational field and matter'' is an old one. In [2], even in
the case of Post-Newtonian approximation, the tension forces have been
accounted by taking into consideration the elastic energy of potential
deformation, and relating this energy to the large-scale translational and
rotational motion of matter in the potential field, created by the
gravitational field. In [3] it has been suggested that the stress tensor is
nothing else, but the ''twice'' projected energy-momentum tensor

\begin{equation}
S_{ik}\equiv p_{il}T_{lm}p_{mk}  \label{2}
\end{equation}
where $p_{il}$ denotes the so-called {projection tensor}

\begin{equation}
p_{il}\equiv \delta _{il}+\frac 1{c^2}u_iu_k  \label{3}
\end{equation}
Since the four-vector $u_{i{ }}$is defined as $u_i\equiv \frac{dx_i}{%
d\tau }$ and $d\tau \equiv \sqrt{1-\frac{v^2}{c^2}}$ is the proper time in
the comoving frame of the particle, $u$ can naturally be assumed to be a
time-like, unit-normalized vector. In [4], the more interesting consequence
from this assumption has been shown - the projection tensor $p_{il}$ can be
regarded as a metric tensor in the subspace, orthogonal to the vector $u$.

It is important to realize that although this is conceived as something very
natural, \emph{the metricity of the projection tensor turns out to play a
crucial role,} and not surprisingly, it is present in many papers, including
most recent ones. For example, in [5] a set of three Lagrangian
space-dimensional coordinates $\xi ^a=\xi ^a(x)$ ($a=1,2,3)$ had been
assigned, and a metric in the three-dimensional ''material'' space is
introduced
\begin{equation}
G^{ab}:=g^{\mu \nu }\xi _\mu ^a\xi _\nu ^b;{ }\mu ,\nu =1,2,3,4
\label{4}
\end{equation}
where $\xi _\mu ^a\equiv \partial _\mu \xi ^a$. The metric $G^{ab}$ enables
one to measure the distances between adjacent particles in the medium. In
fact, the idea about the ''material'' metric is an old one [6] and the
analogy originates from continuous mechanics, where Christoffell symbols
have been introduced to describe the medium's mechanical properties in
curvilinear coordinates. In the case of gravitational theory, this turns out
to be a second metric, besides the initially given space-time metric.
Similar to [4], in [5] it has also been assumed that the velocity vector
field $u$ is a \emph{future oriented, time-like and unit-normalized vector},
orthogonal to $\xi _\mu ^a:$

\begin{equation}
u^\mu \xi _\mu ^a\equiv 0{ \thinspace \thinspace \thinspace \thinspace
\thinspace \thinspace \thinspace \thinspace \thinspace \thinspace \thinspace
\thinspace \thinspace \thinspace \thinspace \thinspace and \thinspace
\thinspace \thinspace \thinspace \thinspace \thinspace \thinspace \thinspace
\thinspace \thinspace \thinspace \thinspace \thinspace \thinspace \thinspace
\thinspace \thinspace \thinspace \thinspace }u^\mu u_\mu =-1  \label{5}
\end{equation}
Note, however that although $\xi _\mu ^a$ is defined on a three-dimensional
(coordinate) subspace, orthogonal to the vector field, it is not yet a
projection tensor. But this might happen if an idempotent endomorphism of
the tangent space onto the same space is defined, according to which

\begin{equation}
p_\beta ^\alpha p_\gamma ^\beta \equiv p_\gamma ^\alpha  \label{6}
\end{equation}
which is of course a typical property of the projection operator and is also
a consequence of the defining equation (3). In the case when the vector
field is not unit-normalized and has a lenght $e=u^\mu u_\mu $, the
projection tensor should be defined as

\begin{equation}
p_{\mu \nu }\equiv g_{\mu \nu }-\frac 1eu_\mu u_\nu  \label{7}
\end{equation}
and the above definition will be used further in the paper.

It is important to mention that the projection tensor definition in [7] and
in the subsequent papers [8-11] is no longer related to any time-like
velocity vector field, unlike the approach for example in [12-13]. In these
papers, in view of tracing the time dependence of the relativistic internal
energy and projecting the energy-momentum conservation condition $\nabla
_\beta T^{\alpha \beta }\equiv 0$ onto a three-dimensional space-like
hypersurface, such an assumption has naturally been preserved. But in a more
general context and from a mathematical point of view [8, 9], the definition
is given in terms of a $(4+n)-$dimensional manifold $M$ with a tensor field $%
\gamma $, assigning a mapping of the tangent space $T_x$ onto the same
space, a set of linearly independent deformation form fields $\theta ^{(A)}$
and also an additional tensor field $H$, called a gauge. The entity $%
(M,\gamma ,\theta ^{(A)})$ is known as a deformation. For all of them the
following conditions are fulfilled for $\gamma $ and $\theta ^{(A)}$:

\begin{equation}
\gamma \theta ^{(A)}\equiv 0  \label{8}
\end{equation}

\begin{equation}
d\theta ^{(A)}\equiv 0  \label{9}
\end{equation}
and the following conditions - for the gauge $H$ (can be identified with the
projection tensor):

\begin{equation}
H_b^aH_c^b\equiv H_c^a{ \thinspace \thinspace \thinspace \thinspace
\thinspace \thinspace \thinspace \thinspace \thinspace \thinspace \thinspace
\thinspace \thinspace \thinspace \thinspace \thinspace \thinspace \thinspace
\thinspace \thinspace \thinspace and \thinspace \thinspace \thinspace
\thinspace \thinspace \thinspace \thinspace \thinspace \thinspace \thinspace
\thinspace \thinspace \thinspace \thinspace \thinspace \thinspace \thinspace
\thinspace \thinspace \thinspace \thinspace \thinspace \thinspace \thinspace
\thinspace \thinspace \thinspace \thinspace \thinspace \thinspace }%
H_{ab}H^{bc}\equiv H_a^c  \label{10}
\end{equation}

\begin{equation}
H_b^a\gamma ^{bc}\equiv \gamma ^{ac}{.}  \label{11}
\end{equation}
More important, however, is that only in a special coordinate chart - the
aligned chart, it is possible to have

\begin{equation}
H_\beta ^\alpha \equiv \delta _\beta ^\alpha {.}  \label{12}
\end{equation}

In other words, if $H_{\alpha \gamma }\equiv p_{\alpha \gamma }$, because of
(10) and (12) the equality

\begin{equation}
p_{\alpha \gamma }p^{\gamma \beta }\equiv p_\alpha ^\beta \equiv H_\alpha
^\beta \equiv \delta _\alpha ^\beta  \label{13}
\end{equation}
will be fulfilled, meaning that only in the particular aligned coordinate
chart the projection tensor has a well-defined inverse one. In the general
case (as can be seen from (7))), an inverse projection tensor will not
exist, but in some special case it can be achieved. A typical example is the
well-known Arnowitt-Deser-Misner (ADM) (3+1) decomposition of space-time
[14, 15]. The ADM\ approach is nothing else but a special kind of a
projection approach, in which identification of components of the vector
field (in ADM\ notations -$N_i$) with certain components of the initial
metric or the projection tensor is made according to the following
substitutions [14, 15] ($i,j=1,2,3)$

\begin{equation}
g_{oo}\equiv -(N^2-N_iN^i){\thinspace \thinspace \thinspace \thinspace
\thinspace \thinspace \thinspace \thinspace \thinspace \thinspace \thinspace
\thinspace \thinspace \thinspace \thinspace \thinspace \thinspace \thinspace
\thinspace \thinspace \thinspace \thinspace \thinspace \thinspace \thinspace
\thinspace \thinspace \thinspace \thinspace \thinspace \thinspace \thinspace
\thinspace \thinspace \thinspace \thinspace \thinspace \thinspace \thinspace
\thinspace \thinspace \thinspace \thinspace \thinspace \thinspace \thinspace
\thinspace \thinspace \thinspace \thinspace \thinspace \thinspace \thinspace
\thinspace \thinspace \thinspace \thinspace \thinspace \thinspace \thinspace
\thinspace \thinspace \thinspace \thinspace \thinspace \thinspace \thinspace
\thinspace \thinspace \thinspace \thinspace \thinspace \thinspace }%
g^{oo}\equiv -\frac 1{N^2}  \label{14}
\end{equation}

\begin{equation}
g_{ij}\equiv
p_{ij}\,\,\,\,\,\,\,\,\,\,\,\,\,\,\,\,\,\,\,\,\,\,\,\,\,\,\,\,\,\,\,\,\,\,\,%
\,\,\,\,\,\,\,\,\,\,\,\,\,\,\,\,\,\,\,\,\,\,\,\,\,\,\,\,\,\,\,\,\,\,\,\,\,\,%
\,\,\,\,\,\,\,\,\,\,\,\,\,\,\,\,\,\,g^{ij}\equiv p^{ij}-\frac{N^iN^j}{N^2}%
\,\,  \label{15}
\end{equation}

\begin{equation}
g_{oi}\equiv
N_i\,\,\,\,\,\,\,\,\,\,\,\,\,\,\,\,\,\,\,\,\,\,\,\,\,\,\,\,\,\,\,\,\,\,\,\,%
\,\,\,\,\,\,\,\,\,\,\,\,\,\,\,\,\,\,\,\,\,\,\,\,\,\,\,\,\,\,\,\,\,\,\,\,\,\,%
\,\,\,\,\,\,\,\,\,\,\,\,\,\,\,\,\,\,\,\,\,\,\,\,\,\,\,\,\,\,\,\,\,\,\,g^{oi}%
\equiv \frac{N^i}{N^2}\,\,  \label{16}
\end{equation}

Since the formulaes are well-known and widely applicable, it is more
important to undestand the advantages of such a substitution, from the
standpoint of a more general projection theory, which is believed that
should exist (for some classical aspects,see also [16, 17]). First, note
that from (15) $p_{ij}p^{jk}\equiv \delta _i^k,$ which follows also from $%
N_iN^i\equiv N^2$ and $N_iN^k\equiv \delta _i^k$. As a result, the
projection tensor has a well defined inverse one. The last fact, although
not commented at all, has been correctly noted also in [18]. Second, from
(15) it follows that if the inilially given metric has zero covariant
derivative, then the projection tensor has also this property, and moreover,
this property is valid for the projection tensor with covariant and
contravariant indices. But it should be stressed that this is a consequence
of the substitutions (14-16), due to which the ADM formalism should be
regarded as a partial case of a more general formalism.

Note also that in the most general case, the definition of a metric (also -
the projection metric) is not related to the existence of an inverse tensor
and in this sense it has a more restricted meaning. For example, within the
class of the so called theories with covariant and contravariant metrics and
affine connections [19], a covariant and a contravariant projection metric
can be defined, in spite of the fact, that an inverse projection tensor may
not exist. Perhaps it should be mentioned that in [20- 22] in an analogous
to (4) way a projection metric on a three dimensional subspace has been
introduced, but with assigned on it space-like Lagrangian coordinates. Of
course, this assumption is needed in this particular case of application of
the formalism of Lagrangian coordinates, but it is not a consequence of the
projection approach in gravitational theory in its most general aspects. As
it can be seen [19], for the introduction of a projection metric, defined by
means of the action of the contraction operator on two vector field from one
and the same vector space, the only assumption that is needed is about the
existence of a non-null vector field, defined over some differentiable
manifold. The applied projectional approach in this paper will use this
definition, allowing a vector field of a more general type. Moreover, the
manifold may be assumed to be of an arbitrary dimension, and the projection
formalism can be therefore generalized for the case of a $p-$ dimensional
submanifold, orthogonal to the complimentary $(n-p)$- submanifold [23]. Such
a model can turn to be useful in multidimensional cosmological models,
including ''brane'' physics. However, this remains out of the scope of the
present paper.

\begin{center}
\bf{II. OUTLINE\ OF\ THE \ METHOD \ IN \ THE \ PRESENT \ PAPER}
\end{center}

The above analyses aims to show the necessity of a more general projection
approach, based on the following assumptions:

1. The vector field $u$ is assumed to be an arbitrary, non-null vector field.

2. The projection tensor does not have an inverse one, which means that as a
result of (7) and the existence of an inverse initial metric tensor $%
g^{\alpha \beta }$, the following relation between the covariant and the
contravariant components of the projection tensor and the vector field is
fulfilled:

\begin{equation}
p_{mk}p^{ik}\equiv \delta _m^i-\frac 1eu^iu_m{.}  \label{17}
\end{equation}
Two more relations shall constitute the basic equations, extensively used
further in the text - the relation, expressing the orthogonality of the
vector field $u$ in respect to the projective tensor, written in the form:

\begin{equation}
\frac 1eu^ku^ip_{km}\equiv 0{,}  \label{18}
\end{equation}

and also the relation for the Riemannian initial metric $g_{ik}$ - zero
covariant derivative $\nabla _\alpha g_{\mu \nu }\equiv 0$ ($\nabla _\alpha
- $ a covariant derivative in respect to the initial Christoffel connection $%
\Gamma _{ij}^s$). It can be expressed as an nonlinear equation between the
vector field $u$, the projective tensor field $p_{ik}$ and their first
partial derivatives:

\begin{equation}
\partial _jg_{ik}\equiv g_{s(k}\Gamma _{i)j}^s  \label{19}
\end{equation}

and a separate equation for the metric tensor with contravariant components:

\begin{equation}
\partial _jg^{ki}\equiv -g^{s(k}\Gamma _{sj}^{i)}{.}  \label{20}
\end{equation}

It follows also that

\begin{equation}
\nabla _\alpha p_{\mu \nu }=-\nabla _\alpha (\frac 1eu_\mu u_\nu )\neq 0%
{.}  \label{21}
\end{equation}
Evidently, this covariant derivative will be zero if $\nabla _\alpha u_\mu
=0 $, but the last would mean that a special kind of transport of the vector
field $u$ has been assumed. Of course, the covariant derivative of $p_{\mu
\nu }$ (denoted by $\widetilde{\nabla { }}$) in respect to the so
called projection connection $\widetilde{\Gamma }_{\alpha \mu }^\gamma $ is
also different from zero, where $\widetilde{\Gamma }_{\alpha \mu }^\gamma $
is defined in the standard way as

\begin{equation}
\widetilde{\Gamma }_{\alpha \mu }^r\equiv \frac 12p^{rs}(\partial _\alpha
p_{\mu s}+\partial _\mu p_{\alpha s}-\partial _rp_{\alpha \mu }){.}
\label{22}
\end{equation}

Note that the projection connection does not have the properties of the
Christoffell connection because it is defined by means of a non-metrical
projection tensor , not having an inverse one.

In [24-preceeding paper] a projection approach has been proposed, and also
some reasoning from a physical point of view for constructing a variational
formalism for a Lagrangian of the kind

\begin{equation}
L=L(p_{mk},p^{mk},\partial _jp_{mk},\partial _jp^{mk},u_{k,{ }%
}u^k,\partial _ju_k,\partial ^ju^k)  \label{23}
\end{equation}
has been given . The Lagrangian (23) is in fact derived from the standard
gravitational Lagrangian, decomposed according to (7). A basic feature of
this projection variational approach is that an account has been taken of
the {form} variations of all vector and tensor quantities with
covariant and contravariant indices and their first partial derivatives. The
approach is similar to that in [25-27], where variations have been taken
also of (generally - non-metric) tensor fields of a mixed type with
covariant and contravariant indices. In the present paper, the role of the
non-metric tensor field is played by the projection tensor and its first
partial derivatives, but another choice is made of taking the variations of
the projection tensor with covariant and separately - with contravariant
components. The choice obviously is dictated by the gravitational Lagrangian
decomposition. Unlike the investigation in [25-27], where the non-metric
tensor fields were not specified or just assumed to be the components of the
affine connection, here the non-metric fields are the projected components $%
p_{\mu \nu }$ of the metric tensor $g_{\mu \nu }$ in respect to the vector
field $u$, which is related to matter and therefore to the reference system.
In [25-27] also variations of the metric tensor (naturally,only with
covariant componets) and its partial derivatives have been accounted . In
the present case, however, variations of $g_{\mu \nu }$ are not to be taken
account, because the variation itself is applied after the Lagrangian
decomposition has been performed. Instead of $g_{\mu \nu }$, variations of
the vector field with covariant and contravariant components will be taken.
From a physical point of view, the inclusion of the vector field $u$ from
the gravitational part of the Lagrangian in the variational approach may
have serious physical implications, at least because there will be an
additional equation of motion for the vector field.

Performing the {form (or functional) variation }means that the
variational operator acts on the Lagrangian action functional just by acting
only on the under-integral expression, and not on the volume element

\[
\overline{\delta }L\equiv \overline{\delta }\left[ \int
L(p_{mk},p^{mk},\partial _jp_{mk},\partial _jp^{mk},u_{k,}{ }%
u^k,\partial _ju_k,\partial ^ju^k){ }d^4x\right] =
\]

\begin{equation}
=\int \overline{\delta }L(p_{mk},.....,\partial ^ju^k)d^4x=0{.}
\label{24}
\end{equation}

This is a peculiar feature of the form - variational operator, which is
understood as the difference between the functional values, taken at one and
the same point.

\begin{equation}
\overline{\delta p}_{ij}\equiv p_{ij}^{^{\prime }}(x)-p_{ij}(x){.}
\label{25}
\end{equation}

This is unlike the the {total variational operator}, defined as
\begin{equation}
\delta u_i(x)\equiv u_i^{^{\prime }}(x^{^{\prime }})-u_i(x){,}
\label{26}
\end{equation}
where the prime sign ''$^{\prime }$ '' means that both the argument $x$ and
the functional values are being varied. The total variational operator acts
in a more special way on the whole action functional and on the volume
element particularly. In spite of the fact that in gravitational theory
mainly the form-variation is applied (for some mathematical aspects - see
[28]), the total variational operator may also play a signifficant role in a
theory, where a unit volume (matter) element is subjected to an expansion,
motion and deformation. Since these physical processes are related to the
introduced in the theory rotation tensor, deformation and expansion tensors
[4], it is clear that in such a ''form non-invariant'' theory after
performing a total variation of the gravitational Lagrangian there will be
an additional contribution (compared with the form-variated Lagrangian),
expressable in terms of the above mentioned variables. An interesting
definition of the form-variation from a purely mathematical point of view as
a sequence of the ''non-commuting'' functional, Lie and covariant variations
is given also in [29]. But since this is a more subtle and complicated
subject, this will be treated in another paper. In this paper, for
convenience the symbol $\delta $ will denote everywhere a form (functional)
variation.

The result of the form-variation in (24) will be the equation

$\overline{\delta }L(p_{mk},.....,\partial ^ju^k)\equiv 0$ (i.e. equal to
zero under-integral expression ),
written as:

\[
\frac{\delta L}{\delta p_{mk}}\delta p_{mk}+\frac{\delta L}{\delta \partial
_jp_{mk}}\delta \partial _jp_{mk}+\frac{\delta L}{\delta p^{mr}}\delta
p^{mr}+\frac{\delta L}{\delta \partial _jp^{mr}}\delta \partial _jp^{mr}+
\]

\begin{equation}
+\frac{\delta L}{\delta u_k}\delta u_k+\frac{\delta L}{\delta \partial _ju_k}%
\delta \partial _ju_k+\frac{\delta L}{\delta u^k}\delta u^k+\frac{\delta L}{%
\delta \partial _ju^k}\delta \partial _ju^k\equiv 0{.}  \label{27}
\end{equation}

In order to write down the corresponding equations of motion for the
independent variables $p_{ik}$ and $u_i$, all variations should be
explicitely written. In the usual variational approach of gravitational
theory, this is trivial , since it is normal to assume that the variation
and the partial derivative commute, provided however that connection
form-variation is zero [24]. But also in [24] it has been proved that in the
case of projection gravitational theory, this is much more complicated since
the variation and the the partial derivative do not commute. The exact
expression for the different from zero commutator $\left[ \delta ,\partial
_j\right] p_{ik}$ has also been found, and it will be used in Section III.

It is the purpose of the present paper to express all the other variations
in (27), and this requires the computation of all the variations not only of
the projection tensor and the vector field with covariant indices, but also
with contravariant ones. This has been performed in Section III, using the
initial set of defining equations (17-18), and also the formulae for the
vector field's length $e=u_ku^k$. Moreover, combining all the variations and
noticing that some of them can be expressed through others, it turned out to
be possible to express all of them only through variations of $\delta u_t$, $%
\delta p_{ik}$ and $\partial _j\delta u_t$ (or $\partial _j\delta u_t$,
which is the same, since $\partial _j$ and $\delta $ commute in respect to
the vector field $u$). The derived expressions will be the starting point
for constructing the adequate variational formalism and finding the
conserved quantities in the subsequent paper.

Motivated from the given in the Introduction reasoning about the importance
to consider the case of zero covariant derivative of the projection tensor
(in respect to the projection connection) and the analogy made with the
three-dimensional projection tensor in the ADM formalism, being defined on a
three dimensional Riemannian subspace of the initially given four
dimensional space, some relations between the variations have been found in
section IV , V and VI under the above particular assumption. Note that there
is no full analogy with the ADM case, because no substitutions like those in
(16-18) have been made. Due to this the result obtained in the end of
section VI is not known from the ADM formalism. In a sense, it may be
admitted that the ADM\ approach admits a more general freedom in the theory,
because the vector field is not assumed to be necessarily orthogonal to the
three-dimensional hypersurface (subspace), as it is required in the
projection approach, used here. The formulaes are valid also for an
arbitary-dimensional space-time, and yet no 3+1 decomposition has been
performed. But in the case of a (3+1) decomposition, the derived formulaes
can be used to check whether the ADM\ formalism can be obtained as a
limiting case of the more general approach, presented so far. This is an
interesting investigation, also not presented in this paper.

Section IV will deal with the variations of the projection tensor with
covariant components, making an extensive use of all the formulaes in
Section III. At the end, a relation between the projection and initial
connections will be obtained.

In Section V the projection connection variation will be found, and the
approach will be based on the fact that in the proposed projection variation
approach different results may be obtained if at first two systems of
equations are combined, and after that the variation is taken, in comparison
with an approach, when first the variations of the two equations are taken,
and only after that the obtained equations are combined. The obtained result
will be possible to be written in two ways. In the second way, no divergent
term will be present but just the variations of the initial connection, the
vector field and the projection tensor.

Section VI, similarly to Section IV, will be devoted to finding the
variations also of the projection tensor, but with contravariant components.
Although the approach will be the same as that in IV, the result about the
expressed projection connection will make it possible to obtain an equation,
expressing the relation between only the vector field and the projection
tensor variations. Since the two variations are independent, the expressions
before them have to be zero, and from one of the expressions an important
relation for the vector field derivative will be obtained in a very concise
form. The significance of the obtained relation will be discussed in the
conclusion part of the paper.

\begin{center}
\bf{III. VARIATIONS \ OF \ FIRST-ORDER \ DERIVATIVES \ OF \ THE \ VECTOR
\ FIELD\ AND\ }

\bf{THE \ PROJECTION \ TENSOR}
\end{center}

In order to implement the projectional variational formalism in the
investigated gravitational theory, using as basic variables the projection
tensor field with covariant and contravariant components, it is necessary to
find all the variations $\delta \partial _ju^k$ and $\delta \partial
_jp^{ik} $ of the first derivatives of the vector field $u^k$ and the
projection tensor $p^{ik}$ with contravariant components. In order to find
the variation $\delta \partial _ju^k$ for example, the commutator $\left[
\delta ,\partial _j\right] $ will be applied to the equation $%
u^kp_{mk}\equiv 0$ and as a result it can be derived

\begin{equation}
p_{mk}\delta \partial _ju^k\equiv -(\partial _ju^k)\delta p_{mk}-(\partial
_jp_{mk})\delta u^k-u^k\delta \partial _jp_{mk}{.}  \label{28}
\end{equation}

Our aim will be to find also an expression for $\frac 1eu_mu_k\delta
\partial _ju^k$, which, if summed up with (28) and subsequently contracted,
will give the required expression for $\delta \partial _ju^k$. This can be
done by applying the commutator $\left[ \delta ,\partial _j\right] $ to the
equation

\begin{equation}
p_{mr}p^{rk}\equiv \delta _m^k-\frac 1eu^ku_m{.}  \label{29}
\end{equation}
The resulting equation, multiplied by $u_k$, is

\begin{equation}
\frac 1eu_mu_k\delta \partial _ju^k\equiv -u_k\delta \partial
_j(p_{mr}p^{rk})-u_k\partial _j(\frac 1eu_m)\delta u^k-e\delta \partial
_j(\frac 1eu_m)-u_k(\partial _ju^k)\delta (\frac 1eu_m){.}  \label{30}
\end{equation}
If (28) and (30) are summed up and contracted with $g^{ms}$, an expression
can be obtained, which contains the other undetermined yet variation $\delta
\partial _jp^{ik}$. In order to avoid this, a more reasonable choice can be
performed by applying the commutator $\left[ \delta ,\partial _j\right] $ to
the equation

\begin{equation}
e=u_ku^k  \label{31}
\end{equation}
Taking into account that $\delta $ and $\partial _j$ commute when applied to
the scalar quantity $e$, i.e.

\begin{equation}
\left[ \delta ,\partial _j\right] e\equiv 0  \label{32}
\end{equation}
the following expression can be obtained (multiplied with $\frac 1eu_m$)

\begin{equation}
\frac 1eu_mu_k\delta \partial _ju^k\equiv -\frac 1eu_mu^k\delta \partial
_ju_k+\frac 1eu_mu^k\partial _j\delta u_k+\frac 1eu_mu_k\partial _j\delta u^k%
{.}  \label{33}
\end{equation}
Now, summing up (28) and (32) and contracting with $g^{ms}$, we obtain the
final expression for $\delta \partial _ju^s$:

\begin{equation}
\delta \partial _ju^s\equiv -g^{ms}u^k\delta \partial
_jp_{mk}-g^{ms}\partial _ju^k\delta p_{mk}-(\partial _jp_{mk})g^{ms}\delta
u^k-\frac 1eu^su^k\delta \partial _ju_k+\frac 1eu^s\partial _j\delta e{.%
}  \label{34}
\end{equation}
\thinspace As can be noted, this expression does not contain variations of
first-order derivatives of the projective tensor with contravariant indices.

Following the same described above procedure and applying the commutator $%
\left[ \delta ,\partial _j\right] $ to the equations

\begin{equation}
p^{mk}u_k\equiv 0{ \thinspace \thinspace \thinspace \thinspace
\thinspace \thinspace \thinspace \thinspace \thinspace \thinspace \thinspace
\thinspace \thinspace \thinspace \thinspace \thinspace \thinspace \thinspace
\thinspace \thinspace \thinspace and \thinspace \thinspace \thinspace
\thinspace \thinspace \thinspace \thinspace \thinspace \thinspace \thinspace
\thinspace \thinspace \thinspace \thinspace \thinspace \thinspace \thinspace
\thinspace \thinspace \thinspace \thinspace \thinspace }p^{mk}p_{ks}\equiv
\delta _s^m-\frac 1eu^mu_s{,}  \label{35}
\end{equation}
an expression for the other variation $\delta \partial _jp^{mr}$ can be
derived:

\[
\delta \partial _jp^{mr}\equiv -\partial _j\left[ \delta (\frac
1e)u^mu^r\right] +\left[ (\partial _ju^r)u^m-g^{sr}u^m\partial _ju_s\right]
\delta (\frac 1e)-
\]

\[
-(\frac 1eu^r\partial _ju_k+g^{sr}\partial _jp_{ks})\delta
p^{mk}-g^{sr}p^{mk}\delta \partial _jp_{ks-}
\]

\[
-g^{sr}\partial _jp^{mk}\delta p_{ks}-g^{sr}\partial _j(\frac 1eu_s)\delta
u^m-\frac 1eu^r\delta \partial _ju^m-
\]

\begin{equation}
-\left[ g^{kr}\partial _j(\frac 1eu_m)+\frac 1eu^r\partial _jp^{mk}\right]
\delta u_k-\frac 1e\left[ u^{(r}g^{m)k}+\frac 1eu^ru^mu^k\right] \delta
\partial _ju_k{.}  \label{36}
\end{equation}
Now it remains to find the variations $\delta p^{mr}$ and $\delta u^k$.
Again, taking the variations of equations (35) and then summing up, it can
be found

\begin{equation}
\delta p^{mr}\equiv -\frac 2eu^rg^{mk}\delta u_k-\frac 1eu^r(g_k^m-\frac
1eu^mu_k)\delta u^k-g^{sr}p^{mk}\delta p_{ks}{.}  \label{37}
\end{equation}

In the same way, after taking the variations of the equations

\begin{equation}
p_{mk}u^k\equiv
0\,\,\,\,\,\,\,\,\,\,\,and\,\,\,\,\,\,\,\,\,\,\,\,\,\,\,e\equiv u_ku^k{,%
}  \label{38}
\end{equation}
the following formulae can be obtained

\begin{equation}
\delta u^k\equiv -u^r(g^{mk}+\frac 1eu^ku^m)\delta p_{mr}\equiv
-u^rg^{mk}\delta p_{mr}{.}  \label{39}
\end{equation}
where the last term in (39) has dissappeared because due to the
orthogonality of the vector field and the projection field the following
relation is fulfilled

\begin{equation}
u^ru^m\delta p_{mr}\equiv 0{.}  \label{40}
\end{equation}

Formulae (39) is convenient since it expresses a variation of a
contravariant quantity {only }through the variation of a covariant
projection tensor. If (39) is substituted into (37) for $\delta p^{mr}$,
again in the right-hand side only variations of covariant quantities will be
present

\begin{equation}
\delta p^{mr}\equiv -\frac 2eu^rg^{mk}\delta u_k-p^{rs}p^{mk}\delta p_{ks}%
{.}  \label{41}
\end{equation}

In the preceeding paper [24] the following expression has been found for the
commutator $\left[ \delta ,\partial _j\right] p_{mr}$

\begin{eqnarray}
\left[ \delta ,\partial _j\right] p_{mr} &\equiv &g_{ir}g_{mk}\partial
_j\left[ \delta p^{ik}+\delta (\frac 1eu^iu^k)\right] +\partial _j\delta
p_{mr}+  \nonumber  \label{42} \\
&&\ \ +g_{k(r}\left[ p^{si}g_{m)i}+\frac 1eu_{m)}u^s\right] \delta \Gamma
_{sj}^k+\frac 1eg_{l(r}u_{m)}\Gamma _{kj}^l\delta u^k+  \nonumber \\
&&\ \ +\Gamma _{sj}^lg_{l(r}\left[ \frac 1eu_ku_{m)}\delta _n^s+\frac
1eu^su_ng_{m)k}\right] \delta p^{nk}-  \nonumber \\
&&\ \ -g_{l(r}g^{np}p_{pm)}\Gamma _{sj}^lg^{sk}\delta p_{nk}-\frac
1{e^2}\Gamma _{nj}^lg_{l(r}u_{m)}u^nu^k\delta u_k{.}  \label{42}
\end{eqnarray}

This formulae contains also variations of the contravariant quantities $%
\delta u^k$ and $\delta p^{nk}$, but one can substitute formulaes (39) and
(41) to find the expression for $\delta \partial _jp_{mr},$ expressed also
through only variations of covariant quantities

\begin{equation}
\delta \partial _jp_{mr}\equiv -\partial _j\left[ \frac 1eu_m(2g_r^t+\frac
1eu^tu_r)\delta u_t\right] +\Gamma _{jk}^sP_{rms}^{tk}\delta
u_t+g_{s(r}g_{m)}^k\delta \Gamma _{kj}^s+\Gamma _{jk}^sG_{srm}^{pkq}\delta
p_{pq},  \label{43}
\end{equation}
where $P_{rms}^{tk}$ is expression (A2) in Appendix A and $G_{srm}^{pkq}$ is
the following expression:

\begin{equation}
G_{rms}^{pkq}\equiv
g^{kp}g_{(m}^qg_{r)s}+g_s^pg_{(r}^kg_{m)}^q-g_{s(r}g_{m)}^pg^{kq}-\frac
12g_{s(r}u_{m)}u^qg^{pk}{.}  \label{44}
\end{equation}
Since the expression for $\delta \partial _jp_{mr}$ enters equation (36) for
$\delta \partial _jp^{mr}$, the last one can easily be found to be

\begin{equation}
\delta \partial _jp^{mr}\equiv W_j^{tmr}\delta u_t+V^{tmr}\partial _j\delta
u_t+Y_j^{pqmr}\delta p_{pq}+Z_p^{rmq}\delta \Gamma _{jq}^p{,}
\label{45}
\end{equation}
where $W_j^{tmr}$, $V^{tmr}$, $Y_j^{pqmr}$ and $Z_p^{rmq}$ are given in
Appendix A - formulaes (A1) and (A3-A5) respectively.

And the last expression to be given is the one for $\delta \partial _ju^m$
(34), which is found to be (after using all the above derived expressions)

\begin{equation}
\delta \partial _ju^m\equiv \partial _j\left[ \frac 3eu^tu^m\delta
u_t\right] -(g^{mr}u_s+g_s^mu^r)\delta \Gamma _{rj}^s+A_j^{pqm}\delta
p_{pq}+B_j^{tm}\delta u_t{,}  \label{46}
\end{equation}
where $A^{pq}$ and $B^t$ in (46) denote the following expressions:

\begin{equation}
A_j^{pqm}\equiv \frac 1eu^m\partial _ju_ru^qp^{pr}+\Gamma _{jr}^s\left[
-g_s^pu^{(r}g^{q)m}+\frac e2g^{pq}(g_s^m+\frac 1eu^mu_s)\right]  \label{47}
\end{equation}

\begin{equation}
B_j^{tm}\equiv \frac 3e(u_pg^{mq}+u^qg_p^m)u^t\Gamma _{qj}^p-\Gamma
_{jq}^pg_p^sg^{mk}u^rP_{rks}^{tq}-\frac 1e\partial _ju^ru^m(2g_r^t+\frac
1eu^tu_r){.}  \label{48}
\end{equation}

\begin{center}
\bf{IV. VARIATIONS\ IN\ THE\ CASE\ OF\ ZERO-COVARIANT\ DERIVATIVE\ OF\
THE\ \ \ PROJECTION\ TENSOR\ \ (WITH\ COVARIANT\ INDICES)}
\end{center}

The assumption about a projection tensor with a zero-covariant derivative in
respect to the projection connection means that besides all the variation
formulaes in the preceeding section, two more conditions are being imposed,
from which again some relations between the variations can be found. The
first condition is for zero covariant derivative of the projection tensor
with {covariant indices }

\begin{equation}
\widetilde{\nabla }_jp_{mr}\equiv 0\equiv \partial _jp_{mr}-p_{s(m}%
\widetilde{\Gamma }_{r)j}^s{,}  \label{49}
\end{equation}
and the second condition is the analogous one, but for the {\
projection tensor field with contravariant indices}

\begin{equation}
\widetilde{\nabla }_jp^{mr}\equiv 0\equiv \partial _jp^{mr}-p^{s(r}%
\widetilde{\Gamma }_{sj}^{m)}{.}  \label{50}
\end{equation}

In this section only the first condition shall be investigated. Note that
while in standard Riemannian geometry (49) will be satisfied if the usual
formulae for the Christoffell connections is used (the two formulaes are a
consequence of one another), here this will not happen, because {the
projection tensor does not have an inverse one}. Consequently, after
formulae (22) for the projection connection $\widetilde{\Gamma }_{rj}^s$ is
substituted into (49), one obtains:

\begin{equation}
\frac 1{2e}u^lu_{(m}\left[ \partial _{r)}p_{jl}+\partial _jp_{r)l}-\partial
_lp_{r)j}\right] \equiv 0,  \label{51}
\end{equation}
which is fulfilled when

\begin{equation}
u^l\left[ \partial _rp_{jl}+\partial _jp_{rl}-\partial _lp_{rj}\right]
\equiv 0{.}  \label{52}
\end{equation}
After taking the variation of this equation, with account also of the
variations (39) and (43), the following equation can be obtained:

\[
\left[ -2u^q\widetilde{\Gamma }_{rj}^p+u^lF_{jlr}^{mn\alpha
}G_{snm}^{pkq}\right] \delta p_{pq}+u^lF_{jlr}^{mn\alpha
}g_{s(m}g_{n)}^k\delta \Gamma _{k\alpha }^s+H_{rj}^t\delta u_t-
\]

\begin{equation}
-\partial _\alpha \left[ \frac 1eu^lF_{jlr}^{mn\alpha }u_m(2g_n^t+\frac
1eu^tu_n)\delta u_t\right] \equiv 0{,}  \label{53}
\end{equation}

where $F_{jlr}^{mn\alpha }$ denotes the following expression:

\begin{equation}
F_{jlr}^{mn\alpha }\equiv g_j^mg_l^ng_r^\alpha +g_r^mg_l^ng_j^\alpha
-g_r^mg_l^ng_j^\alpha ,  \label{54}
\end{equation}
$G_{snm}^{pkq}$ is given by (44) and $H_{rj}^t$ is the expression

\begin{equation}
H_{rj}^t\equiv u^l(\Gamma _{rk}^sP_{ljs}^{tk}+\Gamma
_{jk}^sP_{lrs}^{tk}-\Gamma _{lk}^sP_{jrs}^{tk})+\widetilde{H}_{rj}^t
\label{55}
\end{equation}

\begin{equation}
\widetilde{H}_{rj}^t\equiv \frac 1e\left[ (\partial
_{(r}u^t)u_{j)}+(\partial _{(r}u^{(t}u_{j)}u^{n)}u_n-2(\partial _\alpha
u^\alpha )u_rg_j^t-\frac 1e(\partial _\alpha u^\alpha )u^tu_ru_j\right]
{.}  \label{56}
\end{equation}
Simple calculations give also that

\begin{equation}
u^lF_{jlr}^{mn\alpha }g_{s(m}g_{n)}^k\delta \Gamma _{k\alpha }^s\equiv
2u_s\delta \Gamma _{jr}^s  \label{57}
\end{equation}
and the divergent term (the last term in (53) is found to be

\begin{equation}
S_{rj}\equiv -\partial _\alpha \left[ \frac 1e(-2p_j^tu^\alpha
u_r+3p_r^\alpha u^tu_j+3g_j^\alpha u_ru^t)\delta u_t\right] {.}
\label{58}
\end{equation}

The expression, standing in front of the variation $\delta p_{pq}$ in (53)
(the first term ) can be calculated to be

\begin{equation}
u^q\left[ 2\Gamma _{jr}^p-2\widetilde{\Gamma }_{jr}^p+\frac
12g^{pk}u^l(\Gamma _{kl}^sg_{s(j}u_{r)}-\Gamma _{kr}^sg_{s(l}u_{j)}-\Gamma
_{kj}^sg_{s(l}u_{r)})\right]  \label{59}
\end{equation}
If we assume that the variation and the partial derivative commute in
respect to the metric tensor and therefore $\delta \Gamma _{jr}^s\equiv 0$,
then in view of the independent variations $\delta u_t$ and $\delta p_{pq}$
the expressions (55), (58) and (59), standing in front of them should be
equal to zero. From the last expression, particularly, the projection
connection may be expressed through the full connection and its projections
along the vector field as

\begin{equation}
\widetilde{\Gamma }_{jr}^p\equiv \Gamma _{jr}^p+\frac 14g^{pk}u^l(\Gamma
_{kl}^sg_{s(j}u_{r)}-\Gamma _{kr}^sg_{s(l}u_{j)}-\Gamma
_{kj}^sg_{s(l}u_{r)})\equiv \Gamma _{jr}^p+\frac 12g^{pk}u_ru^l\Gamma
_{k[l}^sg_{j]s}  \label{60}
\end{equation}
Note that the second term is a projected antisymmetric (in the indices $l$
and $j$) expression. The last equation, however, should be viewed only as a
{possible, but not obligatory }choice of the projection connection in
the case of zero-covariant derivative of the projection tensor field and
commutation of the derivative and the variation. The reason for this is
simple: Since the variation $\delta p_{pq}$ does not enter the divergent
term (58) and only the variation $\delta u_t$ enters it, equation (53) can
be considered as a system of differential equations in respect to the
variation $\delta u_t$. This equation will have a solution for arbitrary
variations $\delta p_{pq}$, and therefore there will be no need (60) to be
fullfilled. This will be discussed also in the following section.

Nevertheless, a way can be found to express the projective connection
variation through the initially given one. For that purpose, from the
equation, resulting from the initial Riemannian metric

\begin{equation}
\partial _jg_{mk}-g_{s(m}\Gamma _{k)j}^s\equiv 0{,}  \label{61}
\end{equation}

equation (49) is substracted. As a result, the following equation is obtained

\begin{equation}
\partial _j(\frac 1eu_mu_k)-2g_{sm}H_{kj}\equiv 0{.}  \label{62}
\end{equation}
It is understood that the connection decomposition is
\begin{equation}
\Gamma _{kj}^s\equiv \widetilde{\Gamma }_{kj}^s+H_{kj}^s  \label{63}
\end{equation}
and the variations of the covariant and contravariant metric tensors are

\begin{equation}
\delta g_{pq}\equiv \delta p_{pq}+\frac 2eu_p(p_q^t-\frac 1eu^tu_q)\delta u_t
\label{64}
\end{equation}

\begin{equation}
\delta g^{ms}\equiv -g^{mp}g^{sq}\delta g_{pq}{.}  \label{65}
\end{equation}
Performing the variation of (62), contracting afterwards with g$^{ml{ }%
} $ and making use of (63-65), it can be obtained after some calculations
that

\begin{equation}
\delta \widetilde{\Gamma }_{kj}^s\equiv \delta \Gamma _{kj}^s-\frac
12\partial _j\left[ g^{ms}\delta (\frac 1eu_mu_k)\right] +\frac 12\partial
_j(\frac 1eu_mu_k)g^{mp}g^{sq}\delta p_{sq}+\widetilde{P}^t\delta u_t{,}
\label{66}
\end{equation}
where $\widetilde{P}_k^t$ is the expression

\begin{equation}
\widetilde{P}_k^t\equiv \frac 1e\left[ g^{sq}u^m\partial _j(\frac
1eu_mu_k)-g_k^qu_mg^{l(m}\Gamma _{lj}^{s)}\right] (p_q^t-\frac 1eu^tu_q)%
{.}  \label{67}
\end{equation}

Unlike equation (60) and the connection variation, which can be derived from
it, equation (66) {is always fulfilled,} provided the two starting
assumptions about a Riemannian initial metric and zero-covariant derivative
of the projection tensor are fulfilled.

\begin{center}
\bf{V. FINDING\ THE\ CONNECTION\ VARIATION\ AND\ EXPRESSING\ THE\
DIVERGENT\ TERM}
\end{center}

It would have been very convenient and useful if there is a way to find the
variations $\delta \Gamma _{kj}^s$ and $\delta \widetilde{\Gamma }_{kj}^s$.
For the purpose, the following observation can be made. In (49-52) use is
made first of the equation for the zero-covariant derivative (49), as a
second step - the formulae for the projection connection is substituted and
finally - the variation operator is applied. As a result equation (53) is
obtained. Now, let us proceed in a different way. Let us{\ first}
find the variations of equation (49) and of the defining equation (22 ) for
the projection connection, and as a second step, let us combine the two
obtained (after the variation) equations. The claim which will be made is
that {the result which is to be obtained from the two equations will
be different from the already derived equation (53).} In such a way, the two
independently derived equations may be combined to obtain the final result.

Indeed, the variation of the projection connection, using the defining
formulae (22 ) , is

\begin{equation}
\delta \widetilde{\Gamma }_{\ln }^k\equiv \frac 12\delta p^{kt}(\partial
_np_{lt}+\partial _lp_{nt}-\partial _tp_{\ln })+\frac
12p^{kt}F_{ltn}^{pq\alpha }\delta \partial _\alpha p_{pq}{,}  \label{68}
\end{equation}
where $F_{ltn}^{pq\alpha }$ is the familiar expression (54) and $\delta
p^{kt}$ is written again for convenience

\begin{equation}
\delta p^{kt}\equiv -\frac 2eu^tg^{ks}\delta u_s-p^{ts}p^{kr}\delta p_{sr}%
{.}  \label{69}
\end{equation}
Next, the variation $\delta \partial _\alpha p_{pq}$ is to be expressed
after performing the variation of the equation $\partial _\alpha
p_{pq}-p_{r(p}\widetilde{\Gamma }_{q)\alpha }^r\equiv 0$, as a result of
which

\begin{equation}
\delta \partial _\alpha p_{pq}\equiv \delta p_{r(p}\widetilde{\Gamma }%
_{q)\alpha }^r+p_{r(p}\delta \widetilde{\Gamma }_{q)\alpha }^r{.}
\label{70}
\end{equation}
Substituting the last equation into (68), using (69) and transfering all the
terms with $\delta \widetilde{\Gamma }_{\ln }^{k)}$ in the left-hand side,
it is obtained

\begin{equation}
K_{f\ln }^{kgh}\delta \widetilde{\Gamma }_{gh}^f\equiv N_{\ln }^{kab}\delta
p_{ab}-\frac 1eu^tg^{ks}F_{ltn}^{pq\alpha }(\partial _\alpha p_{pq})\delta
u_s{,}  \label{71}
\end{equation}
where the tensor expressions $N_{\ln }^{kab}$ and $K_{f\ln }^{kgh}$ are
expressed as follows

\begin{equation}
N_{\ln }^{kab}\equiv p^{kt}F_{ltn}^{pq\alpha }g_r^ag_{(p}^b\widetilde{\Gamma
}_{q)\alpha }^r-p^{ta}p^{kb}F_{ltn}^{pq\alpha }\partial _\alpha p_{pq}
\label{72}
\end{equation}

\begin{equation}
K_{f\ln }^{kgh}\equiv g_f^kg_l^gg_n^h-\frac
12p_{f(p}F_{ltn}^{pqh}g_{q)}^gp^{kt}\equiv \frac 12\left[ \frac
1eg_{(l}^gg_{n)}^hu^ku_f+g_{(n}^{[g}p_{l)f}p^{h]k}\right] {.}
\label{73}
\end{equation}
where the square brackets in the contravariant indices mean
antisymmetrization, i.e. $\left[ gh\right] =gh-hg$. Because of the
assumption about symmetric initial and projection connection and the
antisymmetrization of the indices $g$ and $h$ in the second term of (73),
this term will not give contribution to the left-hand side of (71).
Therefore, upon contraction with $u^k$ from (71) it can be found

\begin{equation}
u_s\delta \widetilde{\Gamma }_{kj}^s\equiv -\frac 1eu^mu^tF_{kmj}^{pq\alpha
}\partial _\alpha p_{pq}\delta u_t{.}  \label{74}
\end{equation}
Substituting the last expression into (66), a relation is obtained only
between the (initial) connection variation and the projection and vector
field variations

\[
2u_s\delta \Gamma _{kj}^s\equiv \partial _j\delta u_k-u^qg^{mp}\partial
_j(\frac 1eu_mu_k)\delta p_{pq}-
\]

\begin{equation}
-\left[ \frac 1e(\partial _ju_s)(u_kp^{ts}+u^sg_k^t)+2u_sP_{kj}^{st}+\frac
2eu^mu^tF_{kmj}^{pq\alpha }\partial _\alpha p_{pq}\right] \delta u_t{.}
\label{75}
\end{equation}

The expression for $u_sP_{kj}^{st}$ is easily calculated to be
\begin{equation}
u_sP_{kj}^{st}\equiv \frac 1{e^2}(\partial _ju^m)u_mu^tu_k-\frac
1eu^t\partial _ju_k-\frac 2eu_su^l\Gamma _{lj}^s(p_k^t-\frac 1eu^tu_k){.%
}  \label{76}
\end{equation}
Notice that in deriving (75) use has not been made of the previously derived
equation (54) for the variations. That is why (75) can be substituted into
(53) to obtain

\begin{equation}
K_{rj}^{pq}\delta p_{pq}+O_{rj}^t\delta u_t+divergent\,\,\,\,\,\,\,\,{ }%
terms\equiv 0{,}  \label{77}
\end{equation}
where

\begin{equation}
K_{rj}^{pq}\equiv u^qp^{pk}u_r\left[ u^l\Gamma _{kl}^s-u_s\Gamma
_{kj}^s\right] \equiv p^{pk}u^qu^lu_r(\partial _lg_{kj}-\partial _jg_{kl})
\label{78}
\end{equation}

\begin{equation}
O_{rj}^t\equiv H_{rj}^t-\frac 1e(\partial
_ju_s)(u_rp^{ts}+u^sg_r^t)-2u_sP_{kj}^{st}-\frac 2eu^mu^tF_{kmj}^{pq\alpha
}\partial _\alpha p_{pq}  \label{79}
\end{equation}
and the divergent terms are found to be

\begin{equation}
S_{rj}+\partial _j\delta u_r{.}  \label{80}
\end{equation}

As usual, $S_{rj}$ is given by the expression (58). The obtained equation
(77) contains variations only of the vector and the projective fields, but
also a rather complicated divergent term. However, from (53) and (77) a more
simplified expression can be found, in which the part $S_{rj{ }}$of the
divergent term does not participate

\begin{equation}
2u_s\delta \Gamma _{jr}^s+2u^q(\Gamma _{jr}^p-\widetilde{\Gamma }%
_{jr}^p)\delta p_{pq}+(H_{rj}^t-O_{rj}^t)\delta u_t-\partial _j\delta
u_r\equiv 0{.}  \label{81}
\end{equation}
Unlike (77), here the variation $\delta \Gamma _{jr}^s$ is present, but in
the following section this will turn out to be a useful property.

Now let us assume that the variation and the partial derivative commute in
respect to the initial Riemannian metric (for which $\nabla _\alpha g_{\mu
\nu }\equiv 0$)$.$ Then, according to a proved in [24-preceeding paper]
proposition, this is possible if and only if $\delta \Gamma _{rj}^s\equiv 0$%
, provided also that $\nabla _\alpha (\delta g_{\mu \nu })\equiv 0.$ The
reason is that only upon fulfillment of these conditions the expression for
the commutator $\left[ \delta ,\partial _j\right] g_{\mu \nu }$ will be
equal to zero

\begin{equation}
\left[ \delta ,\partial _\alpha \right] g_{\mu \nu }=\delta (\nabla _\alpha
g_{\mu \nu })-\nabla _\alpha (\delta g_{\mu \nu })-\delta \Gamma _{\mu
\alpha }^\sigma g_{\sigma \nu }-\delta \Gamma _{\nu \alpha }^\sigma
g_{\sigma \mu }=0{.}  \label{82}
\end{equation}

So, if $\delta \Gamma _{rj}^s\equiv 0$ and also the independence of the
variations $\delta u_t$ and $\delta p_{pq}$ are assumed, then from the
expression before $\delta p_{pq}$ in (81) it follows that $\Gamma _{jr}^p=%
\widetilde{\Gamma }_{jr}^p.$ The last is possible only when $u=0$ (then the
remaining in (81) terms will also be zero), but this has to be rejected as
contradictory to the performed projective decomposition.

However, another opportunity exists. Since the partial derivative of the
variation $\delta u_r$ also enters (81), it can be considered as a {%
(matrix) system of four differential equations} in respect to the four
vector $\mathbf{\delta u}$. In matrix notations equation (81) can be written
as

\begin{equation}
\partial _j\mathbf{\delta u\equiv }\widetilde{\mathbf{K}}_j\mathbf{\delta u}+%
\widetilde{\mathbf{T}_j}  \label{83}
\end{equation}
where $\widetilde{\mathbf{K}_j}\equiv K_{rj}^t$ is a 4$\times 4$ matrix ($%
r,t=1,2,3,4)$

\begin{equation}
\begin{array}{cccc}
K_{1j}^1 & K_{1j}^2 & K_{1j}^3 & K_{1j}^4 \\
K_{2j}^1 & K_{2j}^2 & K_{2j}^3 & K_{2j}^4 \\
K_{3j}^1 & K_{3j}^2 & K_{3j}^3 & K_{3j}^4 \\
K_{4j}^1 & K_{4j}^2 & K_{4j}^3 & K_{4j}^4
\end{array}\
\label{84}
\end{equation}
and $K_{rj}^t$ denotes the tensor expression, standing before the variation $%
\delta u_t$ in (81)

\begin{equation}
K_{rj}^t\equiv H_{rj}^t-O_{rj}^t  \label{85}
\end{equation}
This formulae can be expressed from (79), and $\widetilde{\mathbf{T}}_j$
denotes the four-vector (column) $T_{jr}$ (when $j$ is fixed)

\begin{equation}
\widetilde{\mathbf{T}_j}\equiv T_{jr}\equiv 2u_s\delta \Gamma
_{jr}^s+2u^q(\Gamma _{jr}^p-\widetilde{\Gamma }_{jr}^p)\delta p_{pq}{ .}
\label{86}
\end{equation}

Note that the matrix system (83) is for each $j$ and $j=1,2,3,4,$so in fact
there are four such systems. More important, this system is a linear one
since the operation of variation of the four vector $\mathbf{u}$ gives
another four-vector $\mathbf{\delta u,}$ not dependent on $\mathbf{u}$.
Although $\widetilde{\mathbf{K}_j}$ and $\widetilde{\mathbf{T}_j}$ depend on
$\mathbf{u}$ and its derivatives, they are independent of $\mathbf{\delta u}$%
.

The solution of the system (83) for each $j$ can also be written in matrix
notations

\begin{equation}
\mathbf{\delta u}\equiv \int e^{\widetilde{\mathbf{K}_j}}dx_j\left[
const.+\int \widetilde{\mathbf{T}}_je^{-\int \widetilde{\mathbf{K}_j}%
dx_j}dx_j\right] {.}  \label{87}
\end{equation}
The solution is also for each $j$, so the couple of indices in (87) has
nothing to do with summation.

It is important to realize that (87) in fact expresses a complicated
relation between the variations $\delta u_t$, $\delta p_{pq}$ and $\delta
\Gamma _{jr}^s,$and therefore, only in the case of the fulfillment of the
last relation, one may define a covariant derivative of the covariant
projection tensor. But in the general case, when the assumption about
independent variations $\delta u_t$ and $\delta p_{pq}$ is explicitely
present, this is not possible, and one is obliged to investigate the other
case with the contravariant projection tensor. The above formulaes shall not
be further used because use will be made of the expression (81) for the
divergent term $\partial _j\delta u_t.$ However, since there will be no such
(divergent) terms in the other case in the next section, such formulaes will
no longer be written.

\begin{center}
\bf{VI. VARIATIONS \ IN \ THE\ CASE \ OF \ ZERO-COVARIANT \ DERIVATIVE \
OF \ THE\ \ }

\bf{\ PROJECTION \ TENSOR\ \ (WITH \ CONTRAVARIANT \ INDICES)}
\end{center}

In general, the approach in his section will be the same as in the
preceeding section. Some of the formulaes in it will turn out to be related
with the formulaes in the present section.

The defining equation for the zero-covariant derivative of the projective
tensor with contravariant indices is

\begin{equation}
\widetilde{\nabla }_jp^{mr}\equiv \partial _jp^{mr}+p^{s(r}\widetilde{\Gamma
}^{m)}\equiv 0{.}  \label{88}
\end{equation}
Performing the variation of this equation and taking into account
expressions (45) and (41) for the variations $\delta \partial _jp^{mr}$ and $%
\delta p^{sr}$ respectively, the following equation can be obtained

\[
-p^{s(r}\delta \widetilde{\Gamma }_{sj}^{m)}\equiv \partial _j\left[
V^{tmr}\delta u_t\right] +\left[ W_j^{rtm}-\frac 2eu^{(r}g^{st}\widetilde{%
\Gamma }_{sj}^{m)}-\partial _jV^{tmr}\right] \delta u_t+
\]

\begin{equation}
+Z_p^{rmq}\delta \Gamma _{jq}^p+\left[ Y_j^{pqmr}-p^{(rq}p^{sp}\widetilde{%
\Gamma }_{sj}^{m)}\right] \delta p_{pq}{.}  \label{89}
\end{equation}

On the other hand, from the equation for the Riemannian initial metric

\begin{equation}
\nabla _jg^{mr}\equiv \partial _jg^{mr}+g^{s(r}\Gamma _{sj}^{m)}  \label{90}
\end{equation}
after substracting (88) and performing the variation, it can be derived that

\[
\delta \partial _j(\frac 1eu^mu^r)+\delta (\frac 1eu^su^{(r})\Gamma
_{sj}^{m)}+\frac 1eu^su^{(r}\delta \Gamma _{sj}^{m)}+
\]

\begin{equation}
+\delta p^{s(r}(\Gamma _{sj}^{m)}-\widetilde{\Gamma }_{sj}^{m)})+p^{s(r}%
\delta (\Gamma _{sj}^{m)}-\widetilde{\Gamma }_{sj}^{m)})\equiv 0{.}
\label{91}
\end{equation}

Making use of the formulaes for the variations in the previous sections, the
variations of the expressions in the first two terms of (91) can be found.
Omitting some cumbersome transformations, the final form for the transformed
equation (91) can be explicitely given

\[
\partial _j\left[ \frac 5{e^2}u^ru^tu^m\delta u_t\right] +\left[ -\frac
1eu^{(r}g^{m)l}u_s+g_s^{(m}p^{r)l}\right] \delta \Gamma _{lj}^s-
\]

\begin{equation}
-p^{s(r}\delta \widetilde{\Gamma }_{sj}^{m)}+M^{pqmr}\delta
p_{pq}+N^{tmr}\delta u_t\equiv 0{,}  \label{92}
\end{equation}

where $M^{pqmr{ }}$and $N^{mrt{ }}$are expressions (A10) and
(A11), given in Appendix A. So far the two independent equations (89) and
(92) have been obtained and they follow naturally from the assumptions about
the zero covariant derivatives of the metric and the projective tensor.

The term $-p^{s(r}\delta \Gamma _{sj}^{m)}$ from (89) can be substituted
into (92) and in such a way the two equations can be combined. The resulting
equation is

\[
\partial _j\left[ (\frac 5{e^2}u^ru^tu^m+V^{rtm})\delta u_t\right] +\left[
N^{rtm}+W_j^{rtm}-\frac 2eu^{(r}g^{st}\widetilde{\Gamma }_{sj}^{m)}-\partial
_jV^{rtm}\right] \delta u_t+
\]

\begin{equation}
+\left[ M^{pqmr}+Y^{pqmr}-p^{(rq}p^{sp}\widetilde{\Gamma }_{sj}^{m)}\right]
\delta p_{pq}\equiv 0  \label{93}
\end{equation}

Surprisingly, in (93) the variation $\delta \Gamma _{lj}^s$ is absent,
because it has turned out that the expression before $\delta \Gamma _{lj}^s$
is

\begin{equation}
Z_s^{rml}-\frac 1eu^{(r}g^{m)l}u_s+g_s^{(m}p^{r)l}\equiv 0{.}
\label{94}
\end{equation}

The first (divergent) term in (93) can be rewritten, using the expression
(81) for the partial derivative of the variation $\partial _j\delta u_t$.
Further, after this term is substituted in (93), the derived equation is
contracted with $u_ru_m$ and use is made of equation (A6) from Appendix A,
according to which

\begin{equation}
5u^t+u_ru_mV^{tmr}\equiv 0{.}  \label{95}
\end{equation}
Due to (95) some terms in the transformed equation (93) will become zero,
and the expression before the variation $\delta \Gamma _{jl}^s$ will also
turn out to be zero

\begin{equation}
2u_s(5u^l+u_ru_mV^{lmr})\delta \Gamma _{jl}^s\equiv 0{.}  \label{96}
\end{equation}
Therefore, the transformed equation (93) turns out to be without any
divergent term, and moreover, it contains only the variations $\delta u_t$
and $\delta p_{pq}$ and not the variation $\delta \Gamma _{jl}^s$

\[
\delta u_t\left[ W_j^{rtm}u_ru_m-u_ru_m\partial _jV^{tmr}-6\partial _j(\frac
1eu^r)u_ru^t-4\partial _ju^t-\frac 2e(\partial _ju^r)u_ru^t\right] +
\]

\[
+\delta u_t\Gamma _{jq}^p\left[ \frac 6eu_pu^qu^t-2u^\alpha u^\beta P_{\beta
\alpha p}^{tq}-4u_pp^{qt}\right] +
\]

\begin{equation}
+\delta p_{pq}\left[ u_ru_mY^{pqmr}+2(\partial _ju_s)u^qp^{ps}-4\Gamma
_{sj}^lg_l^pu^su^q\right] \equiv 0{.}  \label{97}
\end{equation}

Since the variations $\delta u_t$ and $\delta p_{pq}$ are independent, the
expressions, standing in front of them must be zero. For example, using the
formulae (A8) for the expression $u_ru_mY^{pqmr}$, the equation in front of $%
\delta p_{pq}$ is found to be

\begin{equation}
(\partial _ju^q)u^p-u^q\Gamma
_{sj}^l(3u^sg_l^p+2g^{ps}u_l)+g^{kp}u^q(\partial _ju_k)\equiv 0{.}
\label{98}
\end{equation}
Contracting with $u_pu_q$, the following important relation is obtained

\begin{equation}
\partial _je=5u^su_l\Gamma _{js}^l{.}  \label{99}
\end{equation}
The obtained result can be formulated as follows:{\ in a projection
gravitational theory with zero covariant derivatives of the projection
tensor in respect to the projection connection the partial derivative of the
vector field length }$e${\ can be expressed (up to a number factor,
which as insignificant may be omitted ) as the ''twice'' projected along the
vector field initial connection.} In other words, provided the connection
and the vector field are known, the zero covariant derivative condition sets
up a ''certain law'', according to which the vector field length may change
in space and time. This of course means that under the above assumption for
zero covariant derivative the variation of the vector field length turns out
to be an important characteristic.

\begin{center}
\bf{VII. CONCLUSION}
\end{center}

In this paper a formalism has been developed for finding the variations of
the vector field and the projection tensor with covariant and contravariant
indices, and including only the first and not the second partial
derivatives. An essential feature of the developed approach of non-commuting
variation and partial derivative is that the variations are assumed to
satisfy a system of equations, obtained after the variation of the defining
system of equations - two systems of algebraic equations (35) and two
systems (19-20) of differential equations for the derivative of the metric
tensor with covariant and contravariant indices. Of course, in the last
system the projective decomposition (7) is assumed to be performed. After
finding all the variations and expressing all of them through variations of
the vector field and the projection tensor with covariant indices only, the
case with zero covariant derivative of the projection tensor has been worked
out. This has been motivated by the presented argumentation in the
Introduction for the necessity to compare the applied here more general
approach with the already known and widely used ADM (3+1) projection
approach. As noted, the basic feature of this approach is the existence of a
three-dimensional Riemannian projection metric tensor with a well defined
inverse one. This is of course not characteristic for the more general
approach presented here, but mainly due to this reason the partial and
limiting cases such as the ADM approach deserve particular attention, if the
validity of the more general approach has to be tested.

It should be noted that in the present case, because of the defining
equation (17) $p_{mk}p^{ik}\equiv \delta _m^i-\frac 1eu^iu_m$, the
zero-covariant derivative assumption requires also that the equality

\begin{equation}
\widetilde{\nabla }(\frac 1eu^iu_m){ }\equiv 0  \label{100}
\end{equation}
should be fulfilled. Indeed, equation (100), after contraction with $u_iu^m$%
, gives

\begin{equation}
-\partial _je+(\widetilde{\nabla _j}u^i)u_i+u^m(\widetilde{\nabla }%
_ju_m)\equiv 0{,}  \label{101}
\end{equation}
which of course is always fulfilled, so there is no contradiction.

The applied approach in Sections IV, V\ and VI in the paper for considering
the variations of the projection tensor with covariant and contravariant
indices (under the zero-covariant derivative assumption) clearly
demonstrates how important is to combine and relate the results in the two
cases. In particular, it is interesting to see how the divergent term in
equation (93) in Section VI (for the variations of $p^{ij}$) is expressed by
means of the equation for the variations (81), obtained for the previous
case of variations of the projection tensor with covariant indices. Also,
for some particular cases it is of importance which equations are first
combined and then - variated, and which - first variated, and afterwards -
combined together. The results in the both cases turn out to be different,
but this gives an opportunity to make use of all the derived equations.

The main result in this paper is formulae (99), expressing the derivative of
the vector field's length as a twice projected along the vector field
connection. Now let us write down (99) in another way, using the projective
decomposition (7), i.e. $u^su_l=g_l^se-p_l^se.$ The new obtained formulae is

\begin{equation}
\partial _je=-p_l^s\Gamma _{js}^l{ }e+\Gamma _{sj}^se=P_{sj}^se
\label{102}
\end{equation}
and expressing the fact that the vector field's length is proportional to
the partial derivative of the logarithm of the new connection's trace $%
P_{sj}^s$, i.e.
\begin{equation}
e=\partial _j\ln P_{sj}^s  \label{103}
\end{equation}
where $P_{sj}^s=-p_l^s\Gamma _{sj}^l+\Gamma _{sj}^s.$The first term in
(102), more exactly - the expression before $e$, is the projected with the
projection tensor $p_l^s$ initial connection $\Gamma _{js}^l$, thus
accounting the change of the vector field's length in a reference system,
connected with the vector field $u$. The second term in (102) is a familiar
one from affine differential geometry, where the notion of an equiaffine
connection is introduced [30] - this is the connection, which presumes the
existence of a covariantly constant vector or tensor field (in [30] called a
''n-vector''). Since the last means that the covariant derivative of the
tensor field is zero in respect to this connection, an unit volume element
does not change uder a parallel displacement, performed in a space with an
equiaffine connection. More important, the necessary and sufficient
condition for a connection to be equiaffine is just formulae (103), but in
the sense of the definition in [30] with $P_{sj}^s$ $\equiv \Gamma _{sj}^s$.
Evidently, in the present case $P_{sj}^s$ plays the role of an
''equiaffine'' connection. Note also that when $p_l^s\equiv \delta _l^s$ $%
\equiv g_l^s$ and there is no vector field (see (7 )), from the obtained
relation (102) it follows that $\partial _je=0.$ The equality is fulfilled
also when $e=0,$ so it is seen that the formulae is consistent and not
contradictory.

Although in our case we have used the assumption about zero covariant
derivative, similar to the case of the equiaffine connection, the present
formulae is derived within the formalism of non-commuting variation and
partial derivative in respect to the projection tensor. Therefore, the
derivation of a familiar result, but with some physically justified
modification, within a different variational approach sets up some
interesting questions and problems.

The first problem is that the derived formulae (99) can also be used, after
performing a variation, to find new relations between the variations. Again
an equation for the variations will be obtained, and again from the (equal
to zero) expressions before the variations a new formulae may be obtained
for example for the relation between the projection and the initial
connection. Afterwards, the variation again may be performed and so on,
following the same procedure. In such a way the number of the derived
equations may become infinite. Since the formalism of non-commuting
variation and partial derivative depends on the equations used in the system
for the variations, each time different results will be obtained. That is
why here only the defining equations for the vector field and the projection
tensor are used.

The second problem is whether the usual variational formalism with commuting
$\delta $ and $\partial _j$ is a limiting case of a more general case of
non-commuting ones. Fortunately, there is a simple way to check this, making
use of the found in Section III variations. Also, a 3+1 decomposition should
be performed, and the already mentioned defining system (35) and (19-20)
should be solved also in the 3+1 decomposition approach. Since the
variational formalism in the ADM approach is based on commutativity, it may
be expected that after the substitution of all the relations in the
formulaes for the variations, the commutator $\left[ \delta ,\partial
_j\right] p_{kl}$ will turn out to be zero.

It is worth mentioning also that in classical mechanics long time ago
investigations on the commutativity of the variation and the partial
derivative have also been carried out. For example, in [31] it was proved
that the variation (understood as a space displacement) and the time
derivative commute, provided however that the positions of the material
point after and before the variation should be referred to one and the same
moment of time. Since in a gravitational theory space and time
transformations are related, it is perhaps natural to speak about
non-commutativity of the variation and the partial derivative in some cases.

\begin{center}
\bf{APPENDIX\ A}
\end{center}

In this Appendix the exact expressions for $W_j^{tmr}$, $V_{.}^{tmr}$,$%
Y_j^{pqmr}$ and $Z_p^{rmq}$ will be given. These formulaes are the
expressions, standing before the variations $\delta u_t$, $\partial _j\delta
u_t$, $\delta p_{pq}$ and $\delta \Gamma _{jq}^p$ respectively in formulae
(45).

The exact expression for $W_j^{tmr}$ is

\[
W_j^{tmr}\equiv \frac 2eu^sg^{mt}g^{kr}\partial _jp_{ks}+\frac 1eu_k\partial
_jp^{mk}(2g^{tr}+\frac 1eu^tu^r)-\partial _j(\frac 1eu^mp^{tr})-
\]

\[
-\frac 1eu^tu^r\partial _j(\frac 1eu^m)-g^{ms}(p^{rk}-\frac 1eu^ru^k)\Gamma
_{jq}^pP_{ksp}^{tq}+\frac 3{e^3}u^ru^mu^tu^s(\partial _ju_s)
\]

\begin{equation}
+\frac 2{e^3}u^ru^mu^tu_su_p\partial _jg^{sp}+\frac 1{e^2}u^r(\partial
_ju^{[m})u^{t]}+\frac 2{e^2}u^{[s}g^{t]r}u^m(\partial _ju_s){.}
\label{A1}
\end{equation}

The expression $P_{rms}^{tk}$ in (A1) is

\begin{eqnarray*}
P_{rms}^{tk} &=&\frac 2e(p_r^tp_{ms}u^k+p_r^tp_m^ku_s)+\frac
3{e^2}(p_{ms}u^ku^tu_r+p_m^ku_{.}^tu_ru_s)-\frac 4{e^3}u_mu^tu^ku_ru_s+ \\
&&+\frac{u_m}{e^2}\left[
4p_r^tu^ku_s-4u_su_rp^{kt}-10p_{sr}u^ku^t+2p_s^tu^ku_r+3p_r^ku^tu_s\right] +
\end{eqnarray*}
\begin{equation}
+\frac{u_m}e\left[ -2p_{sr}p^{kt}+p_s^tp_r^k+\frac 3{e^2}u^ku^tu_ru_s\right]
{.}  \label{A2}
\end{equation}

The square brackets $[]$ in (A1) in respect to the tensor indices mean
antisymmetrization, i.e. $[mt]=mt-tm.$

The expressions for $V_{.}^{tmr}$,$Y_j^{pqmr}$ and $Z_p^{rmq}$ are given in
the formulaes (A3 - A5) below

\begin{equation}
V^{tmr}\equiv -\frac 1eu^{(r}\left[ g^{m)t}+\frac 3{2e}u^{m)}u^t\right]
\label{A3}
\end{equation}

\[
Y_j^{pqmr}\equiv -\Gamma _{jn}^l(\frac
1eu^ru^qg^{pn}u^mu_l+g_l^pg^{r(n}g^{q)m})+g^{ms}(p^{rq}p^{kp}-\frac
1eg^{kp}u^ru^q)(\partial _jp_{ks})+
\]

\begin{equation}
+\partial _j(\frac 1eu^ru^q)g^{mp}+\frac 1ep^{mp}p^{kq}u^r\partial _ju_k
\label{A4}
\end{equation}

\begin{equation}
Z_p^{rmq}\equiv -g_p^rp^{mq}-g^{rq}p_p^m+\frac 1eu^rg^{mq}u_p+\frac
1eu^rg_p^mu^q{.}  \label{A5}
\end{equation}
Since the twice projected along the vector field $u$ expressions $%
u_ru_mW_j^{tmr}$, $u_ru_mV_{.}^{tmr}$, $u_ru_mY_j^{pqmr}$ and $%
u_ru_m\partial _jV_{.}^{rmt}$ are also used in the calculations in the
preceeding sections, the exact expressions are also given below.

\begin{equation}
u_ru_mW_j^{rtm}\equiv \frac 3eu^t(\partial _je)-u_s(\partial _jg^{ts})+\frac
2eu^tu^s(\partial _ju_s)-u^su^k\Gamma _{jq}^pP_{ksp}^{tq}  \label{A6}
\end{equation}

\begin{equation}
u_ru_mV^{tmr}\equiv -5u^t  \label{A7}
\end{equation}

\begin{equation}
u_ru_m(\partial _jV^{tmr})\equiv \frac 5eu^tu^r\partial _ju_r-5\partial _ju^t
\label{A8}
\end{equation}

\begin{equation}
u_ru_mY^{pqmr}\equiv -u^q\Gamma _{jn}^l\left[ eg^{pn}+2u^ng_l^p\right]
-u^sg^{kp}u^q\partial _jp_{ks}+(\partial _ju^q)u^p-\frac 1eu^r(\partial
_ju_r)u^qu^p  \label{A9}
\end{equation}
Note that the last term will not give contribution in the expression for $%
u_ru_mY^{pqmr}\delta p_{pq{ }}$ because of the property $u^pu^q\delta
p_{pq}=0.$

Further, the expressions $M^{pqmr}$ and $N^{mrt}$ in (92) can be written as
follows

\[
M^{pqmr}\equiv \frac 2{e^2}(\partial _ju_s)u^qp^{ps}u^mu^r-\partial _j(\frac
1eu^{(r})u^qp^{m)p}+\widetilde{\Gamma }_{sj}^{(m}p^{r)q}+
\]

\begin{equation}
+\Gamma _{sj}^l\left[ g_l^{(m}p^{r)p}g^{qs}-\frac
1eg_l^pu^sg^{q(m}u^{r)}-\frac 1eu^qu^{(r}\left( g_l^pg^{sm)}+\frac
12g^{ps}p_l^{m)}\right) \right]  \label{A10}
\end{equation}

\[
N^{mrt}\equiv \frac 2eu^{(r}g_p^{m)}\widetilde{\Gamma }_{qj}^pg^{qt}-\frac
3eu^tu^{(m}\partial _j(\frac 1eu^{r)})-\frac 4{e^2}(\partial
_ju^t)u^ru^m-\frac 2{e^3}(\partial _ju^s)u_su^tu^ru^m+
\]

\begin{equation}
+\Gamma _{jq}^p\left[ \frac 3{e^2}u^{(r}u_pg^{m)q}u^t-\frac
1eu^{(r}g^{m)k}u^nP_{nkp}^{tq}-\frac 2eu^{(r}g_p^{m)}p^{gt}\right]
\label{A11}
\end{equation}

\begin{center}
\bf{Acknowledgements}
\end{center}

The author is grateful to Dr. S. Manoff from the Institute of Nuclear
Research and Nuclear Energetics of the Bulgarian Academy of Sciences for
useful discussions, critical remarks and advices. The author is grateful
also to his colleagues P. Bojilov and D. Mladenov for useful discussions,
also to Prof. B. M. Barbashov and to Prof. N. A. Chernikov from the
Laboratory for Theoretical Physics at the Joint Institute for Nuclear
Research (Dubna) for their interest towards the investigated problem.

This work has been supported partly by Grant No. F-642 of the National
Science Foundation of Bulgaria.

\begin{center}
\bf{REFERENCES }
\end{center}

1. J. L. Synge, \emph{Relativity: The General Theory} (Noth-Holland
Publishing Company, Amsterdam,1960).

2. V. A. Fock, \emph{Theory of Space, Time and Gravitation} (Moscow, 1961)

3. C. Moller, \emph{The Theory of Relativity} (Claredon Press, Oxford, 1972)

4. S.W. Hawking, G.F.R. Ellis, \emph{The Large Scale Structure of Space-time}%
, (Cambridge University Press, Cambridge, 1973)

5. J. Kijowski, G. Magli, \it{Unconstrained Hamiltonian Formulation of
General Relativity with Thermo-elastic Sources}, \emph{Class.Quant. Grav. }%
\bf{15}, 3891 (1998)

6. A. I. Sedov, \emph{Mechanics of Continuous Medium}, vol.1 and 2, (Nauka,
Moscow, 1973)

7. Robert H. Gowdy, \it{Projection Tensor Hydrodynamics: Generalized
Perfect Fluids}, \emph{Gen. Relat. Grav}. \bf{10}, 431 (1979)

8. R. H. Gowdy, \it{Geometrical Spacetime Perturbation Theory: Regular
First Order Structures,} \emph{J. Math. Phys.} \bf{19}, 2294 (1978)

9. R. H. Gowdy, \it{Geometrical Spacetime Perturbation Theory: Regular
Higher Order Structures,} \emph{J. Math. Phys.} \bf{22}, 988 (1981)

10. R. H. Gowdy, \it{Affine Projection -Tensor Geometry: Decomposing the
Curvature Tensor When the Connection is Arbitrary and the Projection is
Tilted}, \emph{J. Math. Phys.} \bf{35}, 1274 (1994)

11. R. H. Gowdy, \it{Affine Projection-Tensor Geometry: Lie Derivatives
and Isometries},\emph{\ J. Math. Phys.} \bf{36}, 1882 (1995)

12. Gerard A. Maughin, \it{Harmonic Oscillations of Elastic Continua and
Detection of Gravitational Waves}, \emph{Gen. Rel. Grav. }\bf{4}, 241
(1973)

13. G. A. Maughin, \it{On Relativistic Deformable Solids and the
Detection of Gravitational Waves}, \emph{Gen. Rel. Grav.} \bf{5}, 13
(1974)

14. R. Arnowitt, S. Deser, C.W. Misner, \it{The Dynamics of General
Relativity}, in : \emph{Gravitation: An Introduction to Current Research,}
ed. by L. Witten, (John Wiley \& Sons Inc., New York London, 1962)

15. C. W. Misner, K.S. Thorn, J. A. Wheeler, \emph{Gravitation}, vol.2,
(W.H. Freeman and Company, San Francisco, 1973)

16. E. Cartan, \emph{Spaces of Affine, Projection and Conformal Connections}

17. A.P. Norden, \emph{Spaces of Affine Connection}, (Moscow, Nauka, 1976)

18. R. Arnowitt, S. Deser,C. Misner, \it{Dynamical Structure and
Definition of Energy in General Relativity}, \emph{Phys.Rev.} \bf{116},
1322 (1959)

19. S. S. Manoff, \it{Spaces with Contravariant and Covariant Affine
Connections and Metrics}, to appear in \emph{Elem. Part. Atom. Nuclei}
\bf{30}, 1211 (1999)

20. J. D. Brown, K.V. Kuchar, \it{Dust as a Standard of Space and Time
in Canonical Quantum Gravity}, \emph{Phys. Rev}. \bf{D51}, 5600 (1995)

21. K. V. Kuchar, C.G. Torre, \it{Gaussian Reference Fluid and
Interpretation of Quantum Geometrodynamics}, \emph{Phys. Rev}. \bf{D43},
419 (1991)

22. J. D. Brown, D. Marolf, \it{Relativistic Material Reference System},
\emph{Phys. Rev}. \bf{D53}, 1835 (1996)

23. Elias Zafiris, \it{Kinematical Approach to Brane Worldsheet
Deformations in Spacetime}, \emph{Annals of Physics} \bf{264}, 75 (1996)

24. B. G. Dimitrov, \it{A Modified Variational Principle in Relativistic
Hydrodynamics: I. Commutativity and Noncommutativity of the Variation
Operator with the Partial Derivative }(preceeding paper), LANL archive
\bf{gr-qc/9908032}

25. S. S. Manoff, in \it{Lagrangian Formalism for Tensor Fields}, in
\emph{Topics in Comples Analyses, Differential Geometry and Mathematical
Physics}, ed. by S. Dimiev, K. Sekigawa, World Scientific, 1997

26. S. S. Manoff, E\it{instein's Theory of Gravitation as a Lagrangian
Theory for Tensor Fields}, \emph{Intern. Journ. Mod. Phys. }\bf{A13},
1941 (1998)

27. S. S. Manoff, A. Kolarov, B. G. Dimitrov, ($\overline{L}_n,g)-$\it{%
spaces. General Relativity over \ }$\overline{V}_4$\it{\ spaces}, \emph{%
Comm. Joint Inst. Nucl. Res. (Dubna)} \bf{E5-98-184}, 1998

28. I. M. Anderson, \it{Natural Variational Principles on Riemannian
Manifolds}, \emph{Annals of Mathematics} \bf{120}, 329 (1984)

29. S. S. Manoff, \it{Lagrangian Theory of Tensor Fields Over Spaces
With Contravariant and Covariant Affine Connections and Metrics and Its
Application to Einstein Theory of Gravitation in }$\overline{V}_4$\it{\
Spaces, }\emph{Acta Applicandae Mathematicae} \bf{55}, 51 (1999)

30. P. A. Shirokov, A.P. Shirokov, \emph{Affine Differential Geometry},
(Moscow, 1959) (in Russian)

31. A. I. Lurie, Analytical Mechanics, (Moscow, 1961) (in Russian)

\end{document}